\documentclass[%
superscriptaddress,
amsmath,
amssymb,
aps,
onecolumn,
prl,
]{revtex4-2}

\usepackage{graphicx}
\usepackage{dcolumn}
\usepackage{bm}
\usepackage{notes2bib}
\usepackage{trackchanges}
\addeditor{FW}
\usepackage{setspace}
\usepackage{caption}
\usepackage{lineno}

\begin{document}


\title{Amplified Light Beam Cooling via Emergent Onsager's Irreversible Thermodynamics}

\author{Zhongfei Xiong}
\affiliation{School of Optical and Electronic Information, Huazhong University of Science and Technology, Wuhan 430074, China}

\author{Fan O. Wu}
\affiliation{School of Applied and Engineering Physics, Cornell University, Ithaca, NY 14853, USA}

\author{Yang Liu}
\affiliation{School of Physical Science and Technology, Soochow University, Suzhou 215006, China}

\author{Jian-Hua Jiang}
\email{jianhuajiang@suda.edu.cn}
\affiliation{School of Physical Science and Technology, Soochow University, Suzhou 215006, China}
\affiliation{Suzhou Institute for Advanced Research, University of Science and Technology of China, Suzhou 215123, China}

\author{Demetrios N. Christodoulides}
\email{demetri@usc.edu}
\affiliation{Department of Electrical Engineering, University of Southern California, Los Angeles, CA 90089, USA.}

\author{Yuntian Chen}
\email{yuntian@hust.edu.cn}
\affiliation{School of Optical and Electronic Information, Huazhong University of Science and Technology, Wuhan 430074, China}

\affiliation{Wuhan National Laboratory of Optoelectronics, Huazhong University of Science and Technology, Wuhan 430074, China}

\date{\today}
  

\begin{abstract}
High-brightness coherent light source is at the heart of optical technology and yet challenging to achieve. Here, we propose an unconventional approach that utilizes the "forbidden chemical" in optical thermodynamics to convert any incoming light beam into a high-brightness, high-spatial-coherence light beam in multimode nonlinear optical waveguide systems, in contrast to evaporative cooling in cold atoms where the brightness is instead reduced. 
This approach is powered by the fact that light in nonlinear multimode structures undergoes an irreversible thermalization process triggered by its own photon-photon interactions. Moreover, the key characteristics in statistical mechanics, the optical temperature and chemical potential can be widely tuned in photonic systems. As such, when the chemical potential of an optical reservoir is designed to locate at the forbidden band of the probe bosonic system, it can never reach thermal equilibrium with the probe hence endlessly pumping the probe system towards an enhanced brightness and spatial coherence.
This amplified cooling of light beam is revealed via both Onsager’s irreversible thermodynamics theory and numerical simulations. Akin to this effect, the inverse photonic transport currents emerge due to the negative off-diagonal Onsager coefficients. We demonstrate the feasibility of the amplified beam cooling using a coupled multimode optical waveguide system and show that after 800 rounds of amplified cooling, the optical power of an incoming beam is enhanced by 16 times, meanwhile the fundamental mode occupancy is increased to 90\%. These findings unveil an anomalous optical phenomenon and a new route toward high-quality light sources.
\end{abstract}

\maketitle
Recent advances in multi-mode nonlinear optics have shown that a speckled beam can reshape into a ‘clean’ bell-shaped beam in nonlinear multi-mode fiber, termed beam self-cleaning~\cite{krupa2017spatial}. This mechanism 
can be understood and verified as a thermalization phenomenon governed by optical thermodynamic theory (OTT)~\cite{wu2019thermodynamic,parto2019thermodynamic,marques2023observation,makris2020statistical,mangini2022statistical,zhong2023universality,pyrialakos2022thermalization,wu2022thermalization,jung2022thermal,ramos2023theory,wright2022physics}, where photon-photon interaction drives the optical power distribution into equilibrium states following the Rayleigh-Jeans (R-J) distributions (the $i$-mode occupation $ |c_i|^2 = {T}/{\left({\varepsilon _i} - \mu\right)}$) with tunable effective temperature $T$ and chemical potential $\mu$~\cite{wu2019thermodynamic,pourbeyram2022direct,marques2023observation}. 
Similar process has been referred as classical condensations in many previous studies alongside its observation in nonlinear multimode fibers~\cite{aschieri2011condensation,baudin2020classical,baudin2021energy,baudin2023observation}.

\begin{figure}[!ht]
\centering
\includegraphics[width=1\textwidth]{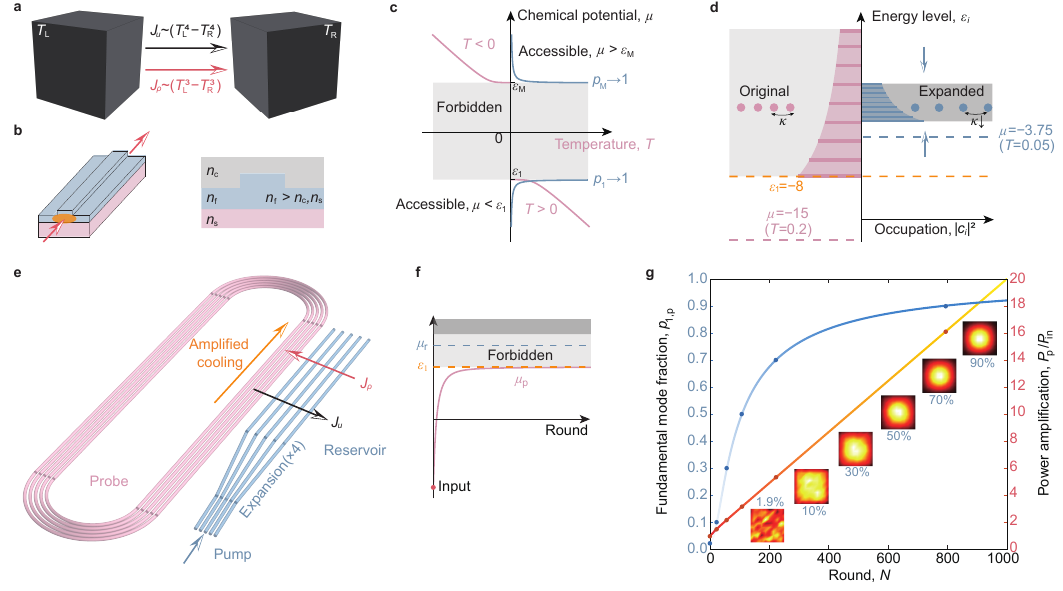}
\captionsetup{font={stretch=1.25},justification=raggedright,singlelinecheck=false}
\caption{\label{IntroductionSketch} \textbf{Forbidden chemical potential and amplified beam cooling.} 
\textbf{a}, In blackbody system, the heat flow ($J_u$) and photon flow ($J_\rho$) is solely proportional to the temperature difference, $\Delta(T^4)$ and $\Delta(T^3)$, respectively. 
\textbf{b}, A ridge waveguide exhibits strong optical confinement due to large refractive index contrast. The refractive indices of the core, cladding, and substrate materials are denoted as $n_{f}$, $n_{c}$, and $n_{s}$, respectively.
\textbf{c}, The effective chemical potential, $\mu$, is forbidden within the spectrum of a waveguide lattice $\varepsilon\in[\varepsilon_1,\varepsilon_M]$ (grey area). As the temperature, $T$, approaches zero from positive/negative values (purple lines), the chemical potential approaches the energy level of the fundamental mode/highest-order mode, $\varepsilon_i$/$\varepsilon_M$, and the fundamental mode/highest-order mode fraction, $p_1$/$p_M$, approaches 1 (blue lines).
\textbf{d}, Altering the waveguide spacing in a nonlinear waveguide array modifies the spectrum, resulting in the expanded system having a forbidden chemical potential relative to the original system. 
The purple and blue bars indicate  the mode occupancies (R-J distributions), $|c_i|^2$, of the original and expanded systems.
The expanded system is obtained by expanding the original system fourfold (reducing the coupling coefficient by a factor of four), while the occupation of the same order mode remains unchanged between the two systems, i.e., the fundamental modes of the two systems have the same occupancy. The chemical potential of the original system (purple dashed line) is lower than the fundamental energy level (orange dashed line), whereas the chemical potential of the expanded system (blue dashed line) is higher than the original fundamental energy level. 
\textbf{e}, The original system (serving as the probe) and the expanded system (serving as the reservoir) constitute an amplified cooling system for light beam. 
The probe is designed as a waveguide loop, and the reservoir is expanded fourfold through a tapered structure.
Inverse power current, $J_\rho$, and energy current, $J_u$, are established between the probe and the reservoir. By repeatedly injecting the incident light beam into the reservoir, continuous ALBC is achieved.
\textbf{f}, The chemical potential of the probe, $\mu_p$, increases and tends towards the fundamental mode energy level of the probe, $\varepsilon_1$, but it cannot reach the chemical potential of the reservoir, $\mu_r$.
\textbf{g}, Fundamental mode fraction and power amplification as a function of the round number. Beam patterns at $p_{1,p}=1.9\%$ (input), $10\%$, $30\%$, $50\%$, $70\%$, and $90\%$, marked by blue and red points, are shown in the inset.}
\end{figure}

The aforementioned classical wave condensation in nonlinear optical systems is promising for developing novel light sources, yet challenged by the limited output power and limited spatial mode purity due to the statistical nature of OTT. One possibility of improving the spatial mode purity, i.e., spatial coherence, is to put two thermal bodies  with orthogonal polarization together, which are coined as the probe and the reservoir~\cite{wu2019thermodynamic,wu2020entropic,ferraro2024calorimetry}.  As the reservoir has a lower temperature than that of the probe, the probe temperature shall decrease  through cross-phase modulation with the reservoir. 
During such nonlinear interactions, the probe beam is cooled down and hence the optical power moves from higher-order modes towards the fundamental mode, leading to an increased spatial coherency. This scheme~\cite{wu2019thermodynamic,wu2020entropic,ferraro2024calorimetry} is limited by the fact that i) the total power in the probe beam will not be amplified during the process, and ii) the cooling efficiency depends on the temperature of the reservoir. At this point, it is natural to ask whether there is a way to simultaneously enhance the power and mode purity of an incoming beam, meanwhile the ultimate cooling efficiency is not limited by Carnot cycles.
Essentially, these require a reservoir that can keep injecting power into the probe beam, meanwhile can constantly extract energy from the probe, leading to an endlessly growing condensation of light.
In contrast to previous studies~\cite{aschieri2011condensation,baudin2020classical,baudin2021energy,baudin2023observation}, throughout this paper, we use the term "condensation" to describe a thermal equilibrium state where almost all the power resides in the fundamental mode ($\gtrsim90\%$).
We remark that in contrast to the conventional near-filed and far-filed electromagnetic radiation where the chemical potential is vanishing~\cite{klaers2010bose} and the transport is solely governed by the temperature, OTT opens a new realm where the transport is controlled by a versatile interplay between temperature and chemical potential --- both can be judiciously tuned in a wide range across positive and negative regimes.
For instance, as illustrated in Fig.~\ref{IntroductionSketch}a, the direction of energy and photon transport between two blackbodies are the same and solely depend on the temperature gradient, in accordance with the Stefan-Boltzmann law, which can never simultaneously enhance the photon number and spatial mode purity of the probe.

Notably, the chemical potential in OTT is essential for increasing the fundamental mode fraction (condensation fraction), which is governed by $p_{1}=\frac{1}{ \left(\varepsilon _1 - \mu \right)}/\left(\sum_{i=1}^{M} {\frac{1}{{\varepsilon _i} - \mu}}\right)$.
In this study, with the help of forbidden chemical potential, we propose a novel yet simple scheme to achieve amplified light beam cooling (ALBC) in nonlinear waveguide lattices.
This is in analogy to tuning the local chemical potential through engineering lattice potential in cold atoms, where cooling can be achieved by  removing atoms located in high potential region~\cite{chiu2018quantum,kantian2018dynamical,yang2020cooling}.  Though engineering refractive index contrast, i.e., similar to the potential of cold atoms, is a routine technique to confine light shown in Fig.~\ref{IntroductionSketch}b in linear optics, its ability to engineer chemical potential explored in cold atoms~\cite{chiu2018quantum,kantian2018dynamical,yang2020cooling} remains elusive in OTT.  In this work, by engineering the effective refractive index (spectrum), we take advantage of  forbidden chemical potential in the reservoir and synthesize inverse current  of internal energy and optical power, which can convert any input beam into an equilibrium state with higher power and mode purity simultaneously.

\hspace{\fill}\\
\noindent\textbf{Forbidden chemical potential}\\
In bosonic systems, for instance nonlinear waveguide lattices, the chemical potential $\mu$ is only permitted outside of the spectrum (below the fundamental mode $\varepsilon_1$ or above the highest-order mode $\varepsilon_M$), as illustrated in Fig.~\ref{IntroductionSketch}c, and is forbidden within the spectrum of supermodes $\left[\varepsilon_1, \varepsilon_M\right]$, in order to ensure a positive value of particle number.
Under extreme conditions, when $T$ approaches $0^+$, $\mu$ approaches $\varepsilon_1$, and the fundamental mode fraction ($p_1$) approaches 1, whereas when $T$ approaches $0^-$, $\mu$ approaches $\varepsilon_M$, and the highest-order mode fraction ($p_M$) approaches 1.
Remarkably, the tunability of optical configurations such as waveguide lattice allows for modifying the forbidden region through spectrum engineering. In this case, engineering mode spectrum is equivalent to engineering the effective refractive index of the supermodes. For instance, increasing the spacing in a nonlinear waveguide array leads to a decrease in the coupling strength $\kappa$ between waveguides, and squeezes the forbidden region of $\mu$. As illustrated in Fig.~\ref{IntroductionSketch}d, when the original system thermalized to $T = 0.2$ and $\mu = -15$ is isentropically expanded by a factor of four ($\kappa \to \kappa/4$), the occupancy of modes remains unchanged, but both $T$ and $\mu$ are reduced by the same factor. Consequently, the new chemical potential $\mu=-3.75$ in the expanded system is now sitting in the forbidden region of the original system. An interesting scenario arises when the new system is strong enough to be considered a reservoir - when the original system is put in contact with such a reservoir, they can never reach thermal equilibrium simply because the chemical potential of the reservoir is forbidden in the original system.

\hspace{\fill}\\
\noindent\textbf{Amplified light beam cooling}\\
To demonstrate ALBC, two subsystems are required. The first is a probe system structured as a waveguide loop, whereas the second system is formed by a straight waveguide that undergoes an isentropic expansion, as indicated by Fig.~\ref{IntroductionSketch}e. The two subsystems are set to have the same initial conditions where the input is an optical beam thermalized to an equilibrium state at $T_{in}=0.2$, $\mu_{in}=-15$, and $p_{1,in}=1.9\%$. Since the second system is constantly pumped, in every round trip, the light in the probe will come back and interact with the same light field in the second waveguide array. Therefore, the second system can be considered a reservoir where the thermodynamic properties remain unchanged during interaction.
As illustrated in Fig.~\ref{IntroductionSketch}d, the isentropic expansion will reduce the coupling strength $\kappa$ of the reservoir by a factor of 4, thus, the temperature and chemical potential in the reservoir becomes $T_{r}=T_{in}/4=0.05$ and $\mu_{r}=\mu_{in}/4=-3.75$ \cite{wu2019thermodynamic,efremidis2021fundamental}.  This, in turn, introduces a temperature gradient and a chemical potential gradient between the two systems, leading to an energy flux and a power flux, as indicated by the fundamental thermodynamic relation. 
It is worth noting that, in optics, even though the internal energy defined as $U=\sum_{i=1}^M{\varepsilon_i|c_i|^2}$ represents the z-component of the electrodynamic momentum flow in the waveguide lattices because it is the conjugate quantity of optical temperature \cite{wu2020entropic}, $U$ and its current are referred to as internal energy and energy flux throughout this paper, 
According to Onsager’s irreversible thermodynamics as detailed in the next section, these gradient
drives energy current $J_u$ from the probe to the reservoir and  inverse photon current $J_\rho$ in the opposite direction, where $J_u$ increases the proportion of the fundamental mode fraction $p_{1,p}$ and photon current $J_\rho$ accounts for the increased optical power in  the probe.

As the incident light is repeatedly injected into the reservoir, as shown in Fig.~\ref{IntroductionSketch}f, the chemical potential of the probe $\mu_p$ shall continuously increase but never be able to reach $\mu_r$ due to the aforementioned forbidden chemical potential of the probe, with respect to the reservoir. Consequently, the chemical potential of probe $\mu_p$ can only approach its ground state energy level $\varepsilon_1$. 
As $\mu_p$ approaches $\varepsilon_1$, the fundamental mode fraction in probe $p_{1,p}$ also approaches 1, which can be termed as beam cooling, as shown in Fig.~\ref{IntroductionSketch}g. Meanwhile, the temperature and chemical potential gradients between the probe and the reservoir also drive the flows, continuously amplifying the probe.
Assuming that the probe and the reservoir reach equilibrium in each round, after nearly 800 rounds, fundamental mode fraction, $p_{1,p}$,  can increase from an initial value of 1.9\% to over 90\%, as shown in Fig.~\ref{IntroductionSketch}g. Concurrently, the total power in the probe is amplified by a factor of 16.

It is crucial to emphasize that the occurrence of inverse current in this scheme is solely caused by the expanded structure of the waveguide lattice, and ALBC possesses a self-driven nature through repeated injection with the same light source. As a result, ALBC is not reliant on the specific temperature of the input light source. 
Compared to beam cooling in ref.~\cite{wu2020entropic,ferraro2024calorimetry}, we use a poor-spatial-coherence reservoir to improve the spatial coherence of the probe, rather than using an extra high-spatial-coherence reservoir to cool the probe.
This approach applies to all positive temperature light sources, and its effectiveness for non-equilibrium light source input is also demonstrated in Supplementary Information. Furthermore, the scheme depicted in Fig.~\ref{IntroductionSketch}e is applicable to light sources with negative temperature and results in both positive currents and an increased occupation of the highest order mode.

To understand the counterintuitive light/heat flow between the probe and reservoir, we first examine the nonequilibrium transport phenomena in nonlinear waveguide lattices, governed by Onsager’s linear irreversible thermodynamics. We then provide a Landauer formula derived from wave turbulence theory (WTT) to determine the direction and magnitude of the flow between the probe and the reservoir.


\hspace{\fill}\\
\noindent\textbf{Onsager's linear irreversible thermodynamics}\\
The fundamental nonequilibrium transport phenomena are the photon flux driven by the density gradient (i.e., the chemical potential gradient) and the energy flow driven by the temperature gradient. The former is governed by Fick's law, while the latter is governed by the generalized Fourier's law, which includes both the positive and negative temperature regimes. The cross-correlated transport is referred as thermoelectric phenomena, specifically the generalized Seebeck effect and the Peltier effect, wherein energy (heat) flow is driven by the temperature gradient and photon flux is driven by the chemical potential gradient.  For simplicity, we discuss these emergent transport phenomena between two optical waveguide lattices with linear coupling shown in Fig.~\ref{Onsager_Current}a (see brief discussion on nonlinear coupled cases in Supplementary Information).
Compared to junction structures \cite{kurnosov2024optical}, the direct coupling between the multimode systems is more feasible to achieve in optics. In practice, two photonic waveguide lattices are linearly coupled by evanescent waves, where the coupling strength between the lattices $\kappa_{ab}$ is much smaller than the coupling strength $\kappa_x$($\kappa_y$) between the waveguides within a lattice. According to the principle of entropy production, Onsager's linear irreversible thermodynamic theory yields the following transport equation (see Methods),
\begin{equation}\label{TwoTerminal}
\bm J = \hat{\bm L} \bm F   
\end{equation}
where $\bm J = [J_{u},  J_{\rho}]^T$, $\bm F= [ F_{u}, F_{\rho}]^T$, and $\hat{\bm L}$ is the Onsager matrix. Explicitly, the energy and photon flows mediated through the interlayer are defined as $J_u=-dU_L/dz=dU_R/dz$ and $J_\rho=-dP_L/dz=dP_R/dz$, respectively, which are driven by the thermodynamic forces given by $F_u=1/T_R-1/T_L$ and $F_\rho=\mu_L/T_L-\mu_R/T_R$, the subscripts L(R) represent the left (right) lattice. We find that Onsager’s reciprocity relation holds~\cite{onsager1931reciprocal}, $\hat{L}=\hat{L}^{\text{T}}$, i.e., $L_{u\rho}$ =$L_{\rho u}$. Alternatively, Eq. \ref{TwoTerminal} can be derived from the WTT (see Methods).

It is a notable check whether the above formalism faithfully describes the transport between photonic lattices, which is an important justification for Eq.~\ref{TwoTerminal}. Figure~\ref{Onsager_Current}b shows that Eq.~\ref{TwoTerminal} indeed agrees with the transport phenomena in the nonlinear photonic lattices.
We perturb two forces $F_u$ and $F_\rho$ separately around the equilibrium state, i.e., $T_L=T_R=1$ and $\mu_L=\mu_R=-20$, and simulate the changes of currents ($J_u$ and $J_\rho$) in response to the forces. The two currents exhibit a linear relationship with the two forces, the slopes of the fitting lines represent the Onsager coefficients at the equilibrium state. Moreover, Onsager's reciprocity relation is revealed by approximately the same slopes of $J_\rho(F_u)$ and $J_u(F_\rho)$, as depicted in Fig.~\ref{Onsager_Current}b. 
As a side remark, the validity of Onsager's reciprocity relation can be determined by examining microscopic reversibility through fluctuation correlation, see details in Supplemental Note 3. We also find that Onsager’s reciprocity applies to all two-terminal thermoelectric systems, which is consistent with previous studies on mesoscopic electronic systems\cite{datta1997electronic,saito2011thermopower,zhu2016persistent,jiang2015efficiency,luo2020onsager}.

The effective thermoelectric effect described in Eq.~\ref{TwoTerminal} and its specific properties can be adjusted by engineering supermode spectrum. For instance, as illustrated in Fig.~\ref{Onsager_Current}c, 
the sign of $L_{u \rho}$ and $L_{\rho u}$ both change in accordance with the on-site detuning $\delta$, which can be controlled by adjusting the waveguide's cross-sectional size or its refractive index. 
At $\delta=0$, $L_{u \rho}=L_{\rho u}<0$, an inverse photon or energy current can occur \cite{wang2020inverse}. In contrast,  both currents are always positive ($L_{\rho u}=L_{u \rho}>0$) regardless of the magnitude of the two positive forces at $\delta<-0.75$. Therefore, the photon and energy transport displays a wide range of inverse currents due to the controllable lattice spectrum of waveguide lattice, even between systems at positive and negative temperatures (Supplementary Note 5). In the following section, we will provide a simplified model to determine two currents and study the occurrence of inverse currents.


\hspace{\fill}\\
\noindent\textbf{Microscopic interpretation of inverse current}\\
From WTT, the photon and energy current can be represented as (see Methods),
\begin{equation}\label{currents}
J_{\rho/u}= \int_{\text{min}}^{\text{max}}{\mathcal{T}(\varepsilon) R_{\rho/u} d\varepsilon}.
\end{equation}
Equation~\ref{currents} conforms to the Landauer formula \cite{kurnosov2024optical} as derived from nonequilibrium Green function method~\cite{saito2007fluctuation,agarwalla2012full,li2012generalized,datta2002non,polanco2021nonequilibrium}. The two currents in Eq.~\ref{currents} are composed of two terms: (1) the transmission coefficient $\mathcal{T}=2\pi|\kappa_{ab}|^2\mathcal{K} D_L D_R$, where $\mathcal{K}$ represents the coupling strength between two supermodes with the same eigenvalue in the two subsystems and $D_L$($D_R$) represents the density of states in the left (right) subsystem, and (2) the R-J factor $R_{\rho}=f_{L}-f_{R}$ proportional to the difference of supermode occupancies $f_{L/R}$ between two subsystems. 
As the coupling strength between the two subsystems is small ($\kappa_{ab}\ll \kappa_{x},\kappa_{y}$), the two subsystems maintain quasi-equilibrium states during the transport process (see Supplementary Note 2). Thus, the distribution $f_{L/R}$ can be approximated with the R-J distribution in equilibrium, as given by $f_{L/R}=T_{L/R}/(\varepsilon_{L/R}-\mu_{L/R})$. As a result, the Onsager matrix can be predicted from Eq.~\ref{currents} (see Methods), 
\begin{equation}\label{OnsagerFromCurrents}
L_{uu/u\rho/\rho\rho}=\int_{\text{min}}^{\text{max}}{ \varepsilon^{2/1/0} \mathcal{T}(\varepsilon) f_L f_R d\varepsilon},
\end{equation}
where $L_{u\rho}=L_{\rho u}$.

According to Eq.~\ref{currents} (or \ref{OnsagerFromCurrents}), the magnitude of the flow (or the Onsager coefficient) is influenced by the transmission coefficient, $\mathcal{T}$, while the direction of the flow is dominated by the R-J factor, $R_{\rho/u}$, and the range of the overlap spectrum between two subsystems, $\left[\varepsilon_{min}, \varepsilon_{max}\right]$. 
Consequently, spectrum engineering can modify $\mathcal{T}$ to change the magnitude of the flow or adjust the range of $\left[\varepsilon_{min}, \varepsilon_{max}\right]$ to alter the direction of the flow.
For example, the lines in Fig. \ref{Onsager_Current}c represent predictions from Eq. \ref{OnsagerFromCurrents}, where the oscillations in the magnitude are primarily attributed to the transmission spectrum (see Supplementary Note 4). Additionally, detuning between the two subsystems reverses the direction of the flow, as demonstrated in Figs. \ref{Onsager_Current}d and \ref{Onsager_Current}e. Figure~\ref{Onsager_Current}d displays $f_{L/R}$ at $F_u=1$, $F_\rho=0.3$ and $\delta = 0$, both the left and right subsystems share an identical spectrum ranging from $\varepsilon_{min}$ to $\varepsilon_{max}$. The R-J factor for photon current $R_\rho$ as shown in the green dashed line is predominantly negative, since the shaded area of positive $R_\rho$ is smaller than that of the negative part. In contrast, the R-J factor of energy current $R_{u}$ is almost always larger than 0 across the entire spectrum, thus energy current is positive. At $\delta=-1$, the spectrum of the left subsystem is modified, ranging from $\varepsilon_{min}'$ to $\varepsilon_{max}'$, as shown in Fig.~\ref{Onsager_Current}e.
As a result, the overlapping spectrum between the two subsystems is truncated to the range from  $\varepsilon_{min}'$ to $\varepsilon_{max}$. In contrast to zero spectrum shift ($\delta = 0$), the spectrum truncation at $\delta=-1$ leads to a significant portion of $R_{\rho}$ being larger than 0, which further changes the direction of the photon current.

\begin{figure}[!ht]
\centering
\includegraphics[width=1\textwidth]{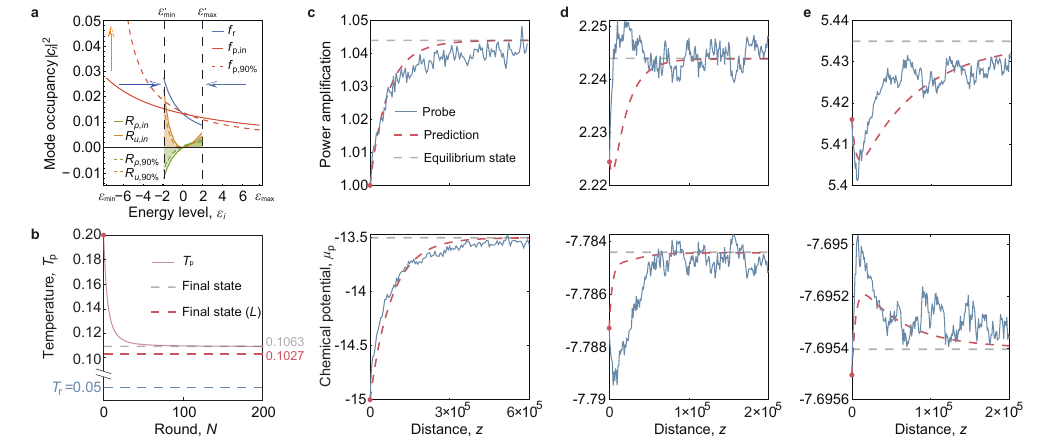}
\captionsetup{font={stretch=1.25},justification=raggedright,singlelinecheck=false}
\caption{\label{AmplifiedCoolingSimulation}
\textbf{Simulation results of amplified light beam cooling.} \textbf{a}, The R-J distributions and R-J factors in Fig.~\ref{IntroductionSketch}e. The red dashed line represents the R-J distribution in $N=796$ round with $p_{1,p}>90\%$. The two systems are both $10\times10$ waveguide lattices. The linear coupling strength $\kappa_x=\kappa_y$ in the probe/reservoir is 2/0.5, $\gamma_{SPM}=-1$ and $\kappa_{ab}=0.1$.
\textbf{b}, Temperature versus round number in the probe. Grey and red dashed lines are predicted asymptotes. The blue dashed line shows the temperature of the reservoir.
\textbf{c,d,e} The simulation results of power amplification $P_p/P_{in}$ and chemical potential $\mu_p$ in the probe at input ($N=0$) with $p_{1,p}=1.9\%$ (\textbf{c}), and $N=56$ with $p_{1,p}=30\%$ (\textbf{d}), $N=222$ with $p_{1,p}=70\%$ (\textbf{e}) in Fig.~\ref{IntroductionSketch}g. The blue solid lines display the simulation results, grey and red dashed lines represent the equilibrium value and the theoretical prediction from Eq.~\ref{Practice} ($\sigma=0.0115$ in \textbf{c} and \textbf{e},  $\sigma=0.0258$ in \textbf{d}), respectively. All simulation results are averaged in 1000 ensemble copies. }
\end{figure}

\hspace{\fill}\\
\noindent\textbf{Transport in amplified light beam cooling}\\
The red/blue line in Fig.~\ref{AmplifiedCoolingSimulation}a shows the R-J distribution $f_p$/$f_r$ of the probe/reservoir under the initial conditions of ALBC in Fig.~\ref{IntroductionSketch}, where $f_r$ is obtained by compressing $f_p$ from the range of $\left[\varepsilon_{min},\varepsilon_{max}\right]$ to the range of $\left[\varepsilon_{min}',\varepsilon_{max}'\right]$. 
As a result, the R-J factor $R_\rho$ implies the presence of an inverse photon flow from reservoir to probe. Similarly, for the R-J factor $R_u$, one can determine the direction of energy flow $J_u$. 
The dashed lines in Fig.~\ref{AmplifiedCoolingSimulation}a display the R-J distribution and R-J factor at round number $N=796$ ($p_{1,p}\approx90\%$), the relation ($J_\rho<0$ and $J_u>0$) still holds at round number $N=796$ and other rounds (not shown). 

As the incident light repeatedly enters the reservoir, the temperature of the probe, $T_p$, does not reach the temperature of the reservoir $T_r$ (represented by the blue dashed line) but approaches a constant value $T_p^{\text{final}}$, as shown in Fig.~\ref{AmplifiedCoolingSimulation}b. 
$T_p^{\text{final}}$ could be predicted when energy current $J_u$ equals $\varepsilon_1 J_\rho$ (see Method),
\begin{equation}\label{TEquil}
T_p^{\text{final}}=\frac{T_r}{M_r} \sum_{i=1}^{M_r}{\frac{\varepsilon_i^r-\varepsilon_1}{\varepsilon_i^r-\mu_r}}.
\end{equation}
Given that the Onsager matrix $\hat{L}$ can be derived from Eq.~\ref{OnsagerFromCurrents}, $T_p^{\text{final}}$ can also be predicted as follows:
\begin{equation}\label{TfinalL}
T_p^{\text{final}}(\hat{L})=\frac{L_{uu}-2L_{u\rho}\varepsilon_1+L_{\rho\rho}\varepsilon_1^2}{L_{uu}-L_{u\rho}(\mu_r+\varepsilon_1)+L_{\rho\rho}\mu_r\varepsilon_1} T_r.  
\end{equation}
Since the probe and reservoir are assumed to reach equilibrium at the end of each round depicted in Fig.~\ref{IntroductionSketch}g, the curve of $T_p$ in Fig.~\ref{AmplifiedCoolingSimulation}b eventually approaches the predicted final temperature $T_p^{\text{final}}$ given by Eq.~\ref{TEquil}, as indicated by the grey dashed line. The red dashed line, predicted from Eq.~\ref{TfinalL} and the initial Onsager matrix, does not coincide with the grey dashed line due to slight changes in the Onsager matrix during propagation. By fixing the length of each round appropriately, we find that $T_p$ approaches the red dashed line rather than the grey dashed line (Supplementary Note 6).

Figures~\ref{AmplifiedCoolingSimulation}c, d and e display the simulated results at the input ($p_{1,p}=1.9\%$, $N=0$), $p_{1,p}=30\%$ ($N=56$) and $p_{1,p}=70\%$ ($N=222$) in blue solid lines, the grey dashed lines represent the equilibrium value for these rounds and the red dashed lines show the prediction result from Eq.~\ref{currents}. In Fig.~\ref{AmplifiedCoolingSimulation}c, the simulation results align with the predictions, validating Eq.~\ref{currents}. The simulation and the predictions do not match well initially in Figs.~\ref{AmplifiedCoolingSimulation}d and e, because under high occupation of the fundamental mode, the SPM of the fundamental mode is too strong, leading to deviations in the predictions obtained from Eq.~\ref{currents} based on WTT. However, as long as the propagation distance is sufficiently long, the photon transport will reach equilibrium. 
According to the fact that the third law of thermodynamics, it is impossible to increase $p_{1,p}$ to 100\% through a finite round of cooling processes~\cite{callen1991thermodynamics,blundell2010concepts}. However, as the condensate fraction $p_{1,p}$ exceeds 70\%, the intensity distribution remarkably resembles the fundamental mode as shown in the inset of Fig.~\ref{IntroductionSketch}f.
As a result, conducting an adequate number of iterations yields a high-brightness and high spatial-coherence light spot representing the fundamental mode of the waveguide array.



\begin{figure}[!ht]
\centering
\includegraphics[width=1\textwidth]{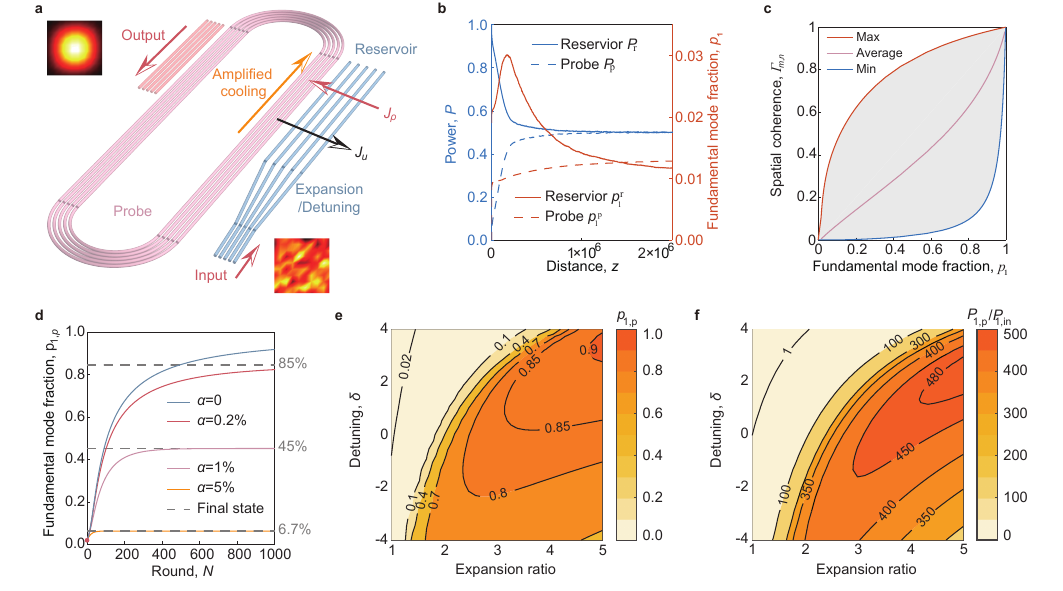}
\captionsetup{font={stretch=1.25},justification=raggedright,singlelinecheck=false}
\caption{\label{SpatialLaser} 
\textbf{High spatial coherence light source based on amplified light beam cooling} \textbf{a}, A schematic of high coherence spatial light source. An output array added in the structure in Fig.~\ref{IntroductionSketch}e and poor beam quality light is pumped in the reservoir. \textbf{b} The simulation results of the variation in power and fundamental mode fraction of the probe and reservoir with the propagation distance  during the initial input ($N=0$, only input in the reservoir). Simulation results are averaged in 1000 ensemble copies. \textbf{c} Spatial coherence increases versus the fundamental mode fraction. The red, purple, and blue lines represent the maximum, average ($\overline{\Gamma_{m,n}}=\sum_{m,n,m\neq n}{\Gamma_{m,n}}$) and minimum value of Eq.~\ref{CrossCorrelation}. \textbf{d}, Given the existence of an output subsystem, the whole system will achieve a steady state within a finite number of rounds. The solid blue line is the curve of the fundamental mode fraction without output (the same line in Fig.~\ref{IntroductionSketch}e). The solid orange, red, and green lines represent the change of fundamental mode fraction with $\alpha=0.002$, $0.01$, and $0.05$, respectively. The dashed grey lines indicate the final steady states. \textbf{e,f}, The contour plots of the proportion and power amplification of the fundamental mode, with the x/y axis representing the reservoir's expansion ratio/detuning. }
\end{figure}

\hspace{\fill}\\
\noindent\textbf{High spatial coherent light source}\\
We proceed to discuss the one possible application of developing novel light sources based on ALBC. 
The reservoir can be seen as an external pump to the gain medium, an effective thermal transport process carries optical power from the pump to the probe with improved mode purity due to ALBC.
Notably, there is no need of see light in probe, as illustrated in Fig.~\ref{SpatialLaser}a. Instead, we only need to input light into the reservoir in the first round, half of the power is coupled into the probe as shown in Fig.~\ref{SpatialLaser}b. 
When the light in the probe and reservoir reach equilibrium, the fundamental mode fraction in the two systems are $p_{1,p}=p_{1,r}=1.3\%$, smaller than the incident light. 
As the number of round trips increases, the fundamental mode fraction rises, resulting in enhanced spatial coherence, similar to Fig.~\ref{IntroductionSketch}g (see Supplementary Note 6).

We can use the cross-correlation function to quantify spatial coherence,
\begin{equation}\label{CrossCorrelation}
    \Gamma_{m,n}(z)=\frac{\langle a_m^*(z) a_n(z) \rangle}{\sqrt{\langle a_m^*(z) a_m(z) \rangle  \langle a_n^*(z) a_n(z) \rangle}},
\end{equation}
where $a_m(z)$ is the complex amplitude of $m$-th site in the probe at certain propagation distance $z$. Fig.~\ref{SpatialLaser}c displays the theoretical relation between the spatial coherence and the condensate fraction $p_1$. 

Finally, we can extract the high spatial coherence light spot through an array that has the same size as the probe, as shown in Fig.~\ref{SpatialLaser}a. This array is weakly coupled to the probe and capable of exporting the beam while preserving an almost unchanged condensate fraction with reduced power (Supplementary Note 6). 
Additionally, if in each round a certain power radio is output through the array, the entire system can eventually reach a steady state. In Fig.~\ref{SpatialLaser}d, we compare the predicted results of three output conditions where $\alpha$ represents the power output radio in each round. The smaller the $\alpha$, the greater the condensate fraction and power amplification at steady state, but more rounds are required. The steady-state labeled by temperature $T_p^s$ and $\mu_p^s$ can be predicted from solve equation (see Method)
\begin{equation}\label{Talpha}
    \sum_{i=1}^{M_r}{\frac{T_p^s}{\varepsilon_i^r P_\alpha-U_\alpha+M_\alpha T_p^s}}+\alpha\sum_{i=1}^{M_p}{\frac{T_p^s}{\varepsilon_i^p P_\alpha-U_\alpha+M_\alpha T_p^s}}=1,
\end{equation}
where $P_\alpha=(1-\alpha)P_r$, $U_\alpha=(1-\alpha)U_r$, $M_\alpha=M_r+\alpha M_p$ and $\mu_p^s=\left(U_\alpha-M_\alpha T_p^s\right)/{P_\alpha}$.
The performance of this scheme is related to the spectrum engineering of the reservoir, as displayed in Figs.~\ref{SpatialLaser}d and f. Positive detuning and larger expansion ratios could increase the fundamental mode fraction up to 90\% and maximize the power gain of the fundamental mode to 480. Under fixed $\alpha$, it is possible to achieve a higher spatial coherent output by adjusting the reservoir. 

\hspace{\fill}\\
\noindent\textbf{Conclusion}\\
We propose a simpler scheme to convert any incoming beam into a single mode by using forbidden chemical potential in nonlinear waveguide lattices. Chemical potential, based on the R-J distribution at equilibrium, is controlled by adjusting the lattice structure, either through coupling strength or on-site detuning. 
Unlike cooling by engineering lattice potential in cold atoms, the negative off-diagonal Onsager coefficient in optical systems supports power amplification while cooling the probe. 
Through the Landauer formula derived from WTT and full discrete nonlinear Schrodinger equations simulations, the amplified cooling between the probe and reservoir can be described via Onsager's irreversible theory.
High power and high fundamental mode fraction in the probe can be achieved by repeatedly injecting the incident light into the reservoir. 
Our work paves the way for developing novel high-brightness, high-spatial-coherent light sources.

The main challenge in ALBC is that modulation instability from the match between the nonlinearity and the diffraction of supermodes~\cite{lederer2008discrete} will hinder ALBC when the ground state occupies high occupancy~\cite{xiong2022k}. We demonstrate in Supplementary Note 6 that $\mu_p$ does not increase at $p_{1,p}=70\%$, in the case of focusing nonlinearity.
Eliminating modulation instability requires the control of the structure's nonlinearity and mode diffraction, such as defocusing nonlinearity (i.e., photorefractive materials~\cite{christodoulides1995bright,nolte2013photorefractive}) for amplified cooling of the fundamental mode, focusing nonlinearity (i.e., Si, GaAs~\cite{agrawal2000nonlinear,boyd2008nonlinear}) for amplified heating of the highest-order mode, or dispersion management~\cite{eisenberg2000diffraction,ablowitz2001discrete}.

\section{Methods}
\noindent\textbf{Nonlinear photonic waveguide lattice and simulation strategy} The (2+1) dimensional waveguide lattice is described by discrete nonlinear Schrodinger equations given from coupled mode theory as follows (see more details in Supplementary Note 1)
\begin{equation}
\label{DNLS}
i \frac{d a_m}{d z}+\delta_m a_m+\sum_{n\neq m}{\kappa_n a_n}+\gamma_{SPM} \left|a_{m} \right|^2 a_{m}=0,
\end{equation}
where $a_m$ is the amplitude of the optical mode field in the waveguide indexed by subscript $m$, $\delta_m$ is the on-site detuning, $n$ represents the label of the nearest waveguides of $m$-th waveguide, $\kappa_n$ is the coupling constant between nearest waveguides, $\gamma_{SPM}$ describes the strength of self-phase modulation in each waveguide as a result of the third-order Kerr effect and $z$ is the normalized propagating distance as pseudo-time. 
The linear part of Eq.~\ref{DNLS} can be expressed as $i d{|\Psi\rangle}/{dz}+\hat{H} |\Psi\rangle=0$, where $|\Psi\rangle=\left(a_1,a_2,...,a_M\right)^\text{T}$ and $\hat{H}$ is the linear Hamiltonian matrix of the waveguide lattice. 
The eigenstates of $\hat{H}$ are a set of supermodes $\psi_i$ with eigenvalue $\beta_i$, and the normalized complex amplitude of the $i$-th supermode is represented by $c_i$.
The eigenvalue $\beta_i$ is the propagation constant relative to the propagation constant $\beta_0$ of the single mode within each waveguide in the array. $\beta_1\geq\beta_2\geq...\geq\beta_{M-1}\geq\beta_M$, where subscript 1($M$) represents the fundamental mode (highest order mode) and $M$ is the number of supermodes. 
We define energy level as $\varepsilon_i=-\beta_i$, then, the fundamental mode has the lowest energy level $\varepsilon_{min}=\varepsilon_1=-\beta_1$ and the highest order mode has the highest energy level $\varepsilon_{max}=\varepsilon_M=-\beta_{M}$. All the simulation results presented in this text are obtained by integrating Eq.~\ref{DNLS} for the entire system.

\noindent\textbf{Transport equation between currents and thermodynamic forces} 
The entropy production of the whole system reads 
\begin{equation*}
\frac{d S_{Total}}{d z}=\sum_{i=L,R}{\frac{d S_i}{dz}}=\sum_{i=L,R}{\frac{1}{T_i}\left(\frac{d U_i}{dz} - \mu_i \frac{d P_i}{dz}\right)}\geq 0.
\end{equation*}
Since the total internal energy $U_{Total}=\sum_{i=L,R}{U_i}$ and the total photon power $P_{Total}=\sum_{i=L,R}{P_i}$ are both conserved, $U_{R}=-U_L$ and $P_R=-{P_L}$. As a result, the entropy production is reduced as $d S_{Total}/{d z}=J_u F_u+J_\rho F_\rho\geq 0$.
As the currents are driven by the forces, each current could be seen as a function of forces. Since two subsystems are close to equilibrium during the process (Supplementary Note 2), one could just use the linear assumption between flows and forces described by Onsager matrix $\hat{L}$, that is
\begin{equation*}
\begin{pmatrix}
J_{u} \\ J_{\rho}   
\end{pmatrix} 
= \hat{L} \begin{pmatrix}
F_{u} \\F_{\rho}  
\end{pmatrix}=
\begin{pmatrix}
L_{uu} & L_{u \rho}  \\
L_{\rho u} & L_{\rho \rho} 
\end{pmatrix} \begin{pmatrix}
F_{u}  \\ F_{\rho}   
\end{pmatrix}.
\end{equation*}

\noindent\textbf{Currents formulation in linear-coupling condition} We consider two lattices $\{a_m\}$ and $\{b_{m'}\}$ coupled in linear condition (see nonlinear condition in Supplementary Note 4).
Firstly, according to Eq. \ref{DNLS}, the dynamic equations for the waveguide without external coupling are written as
\begin{equation*}
\begin{split}
&i \frac{da_m}{dz}+\sum_{n\neq m}{\kappa_a a_n}+\gamma_a \lvert a_{m} \rvert ^2 a_m + \kappa_{a b} b_{h'} \delta_{m,h}=0, \\
&i \frac{db_{m'}}{dz}+\sum_{n'\neq m'}{\kappa_b b_{n'}}+\gamma_b \lvert b_{m'} \rvert ^2 b_{m'} + \kappa_{b a} a_h \delta_{m',h'}=0,
\end{split}
\end{equation*}
where subscript $m$ and $m'$ labels the waveguides in subsystems a and b, respectively, the subscript $h$ ($h'$) labels the waveguides with external coupling (boundary-sites) in each subsystem, the coupling strength between two subsystems is  $\kappa_{a b}=\kappa_{b a}^*$, $\delta_{i,j}$ represents Kronecker delta function.
Secondly, by transforming the above real-space equations into supermode space, we get the dynamic equations for supermodes
\begin{equation*}
\begin{split}
&i \frac{d c_{i}^a}{dz}+\beta_{i}^a c_{i}^a =-\gamma_a\! \sum_{j,k,l}{\Gamma_{i j k l}^a c_{j}^{a*} c_{k}^a c_{l}^a}-\kappa_{ab}\! \sum_{j} {K_{ij}^{ab}c_{j}^b}, \\
&i \frac{d c_{i}^b}{dz}+\beta_{i}^b c_{i}^b =-\gamma_b \sum_{j,k,l}{\Gamma_{i j k l}^b c_{j}^{b*} c_{k}^b c_{l}^b}-
\kappa_{b a} \sum_{j} {K_{ij}^{ba}c_{j}^a}, 
\end{split}
\end{equation*}
where ${c_{i}^a}=\sum_{m}\psi_{i}^{a*}\left(m\right)a_{m}$ (${c_{i}^b}=\sum_{m'}\psi_{i}^{b*}\left(m'\right)b_{m'}$) is the complex amplitude of $i$-th supermode, superscript $a$ and $b$ represents two subsystem, respectively. Coefficient $\Gamma_{ijkl}^a=\sum_m{\psi_{i}^{a*} \psi_{j}^{a*}\psi_{k}^a\psi_{l}^a}$ ($\Gamma_{i j k l}^b=\sum_{m'}{\psi_{i}^{b*} \psi_{j}^{b*} \psi_{k}^b \psi_{l}^b}$) is the overlap summation of four-supermodes in the same subsystem. Coefficient $K_{ij}^{ab}=\sum_{h(h')}{\psi_{i}^{a*} (h) \psi_{j}^b (h')}$ is the overlap summation of two supermodes on boundary-sites $h(h')$ between two subsystems, describing the coupling strength between supermodes $\psi_{i}^{a}$ and $\psi_{j}^{b}$.
Thus, the kinetic equations for the mode occupancies are
\begin{equation*}
\begin{split}
\frac{d n_{i}^a}{dz}=&-2 \text{Im}\!\left(\gamma_a \!\sum_{j,k,l} {\Gamma_{ijkl}^a  c_{ijkl}^{a^*a^*aa} }+\kappa_{a b}\! \sum_{j} {K_{ij}^{ab} c_{ij}^{a^*b}} \right),\\
\frac{d n_{i}^b}{dz}=&-2 \text{Im}\!\left(\gamma_b \sum_{j,k,l} {\Gamma_{ijkl}^b  c_{ijkl}^{b^*b^*bb}}+\kappa_{b a}\sum_{j}{K_{ij}^{ba} c_{ij}^{b^*a}} \right),
\end{split}
\end{equation*}
where $n_{i}^a=\langle c_{i}^{a*} c_{i}^a  \rangle$ is the ensemble average occupation number of $i$-th supermode, $\langle \quad \rangle$ represents ensemble-average (correlation), $\text{Im}(\quad)$ represents imaginary part, $c_{ijkl}^{a^*a^*aa}=\langle c_{i}^{a*} c_{j}^{a*} c_{k}^a c_{l}^a \rangle$ and $c_{ij}^{a^*b}=\langle c_{i}^{a*} c_{j}^b \rangle$.
Hence, the photon current can be represented as the negative summation of the change of mode occupancies in subsystem $\{a_m\}$, that is
\begin{equation}
J_{\rho}=-\sum_{i}{\frac{d n_{i}^a}{dz}}=2\text{Im}\!\left(\kappa_{a b} \sum_{i,j} {K_{ij}^{ab} \langle c_{i}^{a*} c_{j}^b\rangle}\right),
\end{equation}
similarly, the internal energy current is 
\begin{equation}
J_{u}=\sum_{i}{\beta_i^a \frac{d n_{i}^a}{dz}}\approx -2\text{Im}\!\left(\!\kappa_{a b} \!\sum_{i,j} {\beta_{i}^a K_{ij}^{ab} \langle c_{i}^{a*} c_{j}^b \rangle}\right).
\end{equation}
Thirdly, we simplify the correlation of $\langle c_{i}^{a*} c_{j}^b \rangle =c_{ij}^{a^*b}$ with the help of WTT~\cite{nazarenko2011wave,dyachenko1992optical}. The dynamics of $c_{u v}^{a^*b}$ reads
\begin{equation*}
\begin{split}
i \frac{d c_{uv}^{a^*b}}{d z}\!&+\!(\beta_{v}^b-\beta_{u}^a)  c_{uv}^{a^*b}=\kappa_{b a}\!\sum_{j}{\left[(K_{u j}^{ab})^* c_{jv}^{b^*b}-K_{v j}^{b a} c_{ju}^{aa^*}\right]} \\
&+\sum_{j,k,l}{\left[\gamma_a \left(\Gamma_{u j k l}^{a}\right)^* c_{jklv}^{aa^*a^*b}-\gamma_b \Gamma_{v j k l}^b c_{jklu}^{b^*bba^*}\right]}.
\end{split}
\end{equation*}
Since each subsystem maintains
internal equilibrium during transportation, the supermodes between two subsystems could be seen as random phase and amplitude wave~\cite{nazarenko2011wave}.
According to the correlation of random waves $\langle c_i^* c_j\rangle=n_i\delta_{i,j}$ and $\langle c_i^* c_j^* c_k c_l\rangle=n_i n_j (\delta_{i,k}\delta_{j,l}+\delta_{i,l}\delta_{j,k})$, we obtain $c_{jklu}^{b^*bba^*} =0$, $c_{jklv}^{aa^*a^*b}=0$, $c_{ju}^{aa^*}=n_{u}^a \delta_{u,j}$ and $c_{jv}^{b*b}=n_{v}^b\delta_{v,j}$. Therefore, we obtain
\begin{equation*}
i \frac{d \langle c_{i}^{a*} c_{j}^b \rangle}{d z}+(\beta_{j}^b-\beta_{i}^a) \langle c_{i}^{a*} c_{j}^b\rangle=\mathcal{A},
\end{equation*}
where $\mathcal{A}\equiv\kappa_{b a}K_{j i}^{b a}(n_{j}^b-n_{i}^a)$. The special solution at $\beta_{j}^b-\beta_{i}^a=0$ is $\langle c_{i}^{a*} c_{j}^b \rangle=-i \mathcal{A} z$, and the general solution at $\beta_{j}^b-\beta_{i}^a\neq 0$ is
$\langle c_{i}^{a*} c_{j}^b \rangle = \mathcal{C} \exp{[i(\beta_{j}^b-\beta_{i}^a)]}z+\mathcal{A}/(\beta_{j}^b-\beta_{i}^a)$, where $\mathcal{C}$ is a constant.
According to WTT~\cite{nazarenko2011wave,dyachenko1992optical}, we only consider the general solution, ignore its first fast-changing term and add a small imaginary part in the denominator of the second term,
$\langle c_{i}^{a*} c_{j}^b \rangle=\mathcal{A}/({\beta_{j}^b-\beta_{i}^a+i\delta})$,
\begin{equation*}
\kappa_{a b} K_{ij}^{a b}\langle c_{i}^{a*} c_{j}^b \rangle=\frac{|\kappa_{a b}|^2 |K_{ij}^{a b}|^2(n_{j}^b-n_{i}^a)}{\beta_{j}^b-\beta_{i}^a+i\delta}.
\end{equation*}
According to $\frac{1}{x+i\epsilon}=P \frac{1}{x}-i\pi \delta(x)$, the imaginary part of the last equation is
\begin{equation}
\text{Im}\left(\kappa_{a b}  K_{ij}^{a b}  \langle c_{i}^{a*} c_{j}^b \rangle \right)
=\frac{1}{2} \mathcal{T}_{ij} \left(n_{i}^a-n_{j}^b\right)\delta(\beta_{j}^b-\beta_{i}^a), 
\end{equation}
where $\mathcal{T}_{ij}=2\pi |\kappa_{a b}|^2 |K_{ij}^{a b}|^2$ represent the transmission coefficient between supermodes $c_{i}^a$ and $c_{j}^b$.
As a result, the photon/energy currents from subsystem $\{a_m\}$ to subsystem $\{b_{m}\}$ are
\begin{equation}
\label{CurrentInMode}
J_{\rho/u}=\!\sum_{i,j}{ \mathcal{T}_{ij}R_{\rho/u}\delta_{\beta_{j}^b,\beta_{i}^a}}=\sum_{i}{\left(\sum_{j \in B_N}{\mathcal{T}_{ij}R_{\rho/u}}\right)},
\end{equation}
where R-J factor $R_{\rho}=n_{i}^a-n_{j}^b$ represents the difference of mode occupancies between two supermodes and $R_u=-\beta_{i}^a R_\rho$. $B_N=\{ \beta_{j}^b=\beta_i^a \}$ is the set of energy-matched supermodes in two subsystems. Eq.~\ref{CurrentInMode} reveals that the current can be expressed as the sum of the transmission coefficients multiplied by the R-J factor of the energy-matched supermodes between two systems.
Since the supermodes of lattice are uniform in k-space with mode spacing $\Delta \bm{k}={\pi^d}/{M}$, where $d$ is the lattice dimension and $M$ is the number of supermodes, we rewrite Eq.~\ref{CurrentInMode} into k-space
\begin{equation*}
\begin{split}
J_{\rho/u}\!&=\!\sum_{\bm{k},\bm{k}'}{ \mathcal{T}_{\bm{k},\bm{k}'}\!R_{\rho/u}\delta_{\beta_{\bm{k}'},\beta_{\bm{k}}}}\!=\!\sum_{\bm{k}}{\left(\sum_{\bm{k}' \in B_N }{\!\mathcal{T}_{\bm{k},\bm{k}'}\!R_{\rho/u}}\!\right)},\\
&\to\int_{0}^{\pi^d}{ \!\int_{0}^{\pi^d}{\!\frac{\mathcal{T}_{\bm{k},\bm{k}'}R_{\rho/u}}{\Delta \bm{k} \Delta \bm{k'}}   \delta\!\left[\beta^b({\bm{k}'})-\beta^a({\bm{k}}) \right]\!d\bm{k}'}d\bm{k}},
\end{split}
\end{equation*}
where $\bm{k}$ ($\bm{k}'$) is wave vector in subsystem $\{a_m\}$ ($\{b_{m'}\}$). When the lattice is large ($M\to \infty$), the summation in k-space can be approximated to integral in k-space, $\lim_{M\to \infty} \sum_{\bm{k}} \to \frac{1}{\Delta \bm{k}} \int_0^{\pi^d}{d{\bm{k}}} $, $\delta\left[\beta^b({\bm{k}'})-\beta^a({\bm{k}}) \right]$ is Dirac delta function that $\lim_{M\to \infty} \delta_{\bm{k}',\bm{k}}= \Delta \bm{k}' \delta(\bm{k}'-\bm{k})$~\cite{mahan2013many}.
In this scenario, the energy level of supermodes can be also seen as continued quantity, and the formulations of two currents are rewritten as
\begin{equation*}
J_{\rho/u}= \int_{\text{max}(\varepsilon_{min}^a,\varepsilon_{min}^b)}^{\text{min}(\varepsilon_{max}^a,\varepsilon_{max}^b)}{ \mathcal{T}(\varepsilon) R_{\rho/u} (\varepsilon)d\varepsilon},
\end{equation*}
where the transmission coefficient in energy level space is written as $\mathcal{T}(\varepsilon)=2\pi |\kappa_{ab}|^2 \mathcal{K}(\varepsilon) D_a(\varepsilon) D_b(\varepsilon)$. $D_a(\varepsilon)=\frac{1}{\Delta \bm{k}} \frac{d \bm{k}}{d \varepsilon}=\frac{d i}{d \varepsilon}$ and $D_b(\varepsilon)=\frac{1}{\Delta \bm{k'}} \frac{d \bm{k'}}{d \varepsilon}=\frac{d i}{d \varepsilon}$ is the density of states in subsystem a and b, respectively. $\mathcal{K}$ represents the average coupling strength between energy-matched supermodes ($\varepsilon^b=\varepsilon^a$).

In practice, the number of modes in the lattice is finite, and the nonlinear phase shift will change $\varepsilon$. Therefore, when comparing with the simulation results, we replace the delta function in Eq.~\ref{CurrentInMode} with a Gaussian function, 
\begin{equation}
\label{Practice}
J_{\rho/u}= \sum_{i,j}{ \mathcal{T}_{ij}R_{\rho/u}\exp{\left[-\frac{(\varepsilon_i^a-\varepsilon_j^b)^2}{2 \sigma^2}\right]}},
\end{equation}
where $\sigma$ represents the standard deviation of the Gaussian function.

\noindent\textbf{Onsager matrix from currents formula} As two subsystems are assumed to stay in a quasi-equilibrium state during the transportation, mode occupancies $n_a(\varepsilon)$ can be replaced by R-J distribution, $f_a(\varepsilon)=\frac{T_{a}}{\varepsilon-\mu_{a}}$. Thus, we rewrite Eq.\ref{currents} as
\begin{equation*}
\begin{split}
J_{u/\rho}=&\int_{\text{min}}^{\text{max}}{\varepsilon^{1/0} \mathcal{T}(\varepsilon) \left(\frac{\varepsilon-\mu_{b}}{T_{b}}-\frac{\varepsilon-\mu_{a}}{T_{a}}\right) f_a f_b d\varepsilon}\\
=&F_u\int_{\text{min}}^{\text{max}}{\varepsilon^{2/1} \mathcal{T} f_a f_b d\varepsilon}+F_\rho\int_{\text{min}}^{\text{max}}{\varepsilon^{1/0} \mathcal{T} f_a f_b d\varepsilon}.
\end{split}
\end{equation*}
According to Eq.~\ref{TwoTerminal}, we get Eq.~\ref{OnsagerFromCurrents} and find $L_{u\rho}=L_{\rho u}$.
In addition, since $2\varepsilon_i \varepsilon_j \leq \varepsilon_i^2+\varepsilon_j^2$, it's easy to demonstrate that
$(L_{u\rho}+L_{\rho u})^2\leq 4L_{uu}L_{\rho\rho}$
which fulfills the constraint of the transport coefficients imposed by the second law of thermodynamics~\cite{jiang2015efficiency}.

\noindent\textbf{Prediction of final non-equilibrium steady state}  As round number $N\to\infty$ in Fig.~\ref{IntroductionSketch}e, $\mu_p\to\varepsilon_1$, the internal energy flow and power flow satisfy $J_u-\varepsilon_1 J_\rho=0$.
When the propagation distance is sufficiently long for the probe and the reservoir to reach equilibrium in each round, $T_p^{\text{final}}$ satisfies $\Delta U_r/\Delta P_r=\varepsilon_1$, where $\Delta U_r$ and $\Delta P_r$ are the change of internal energy and power in the reservoir, respectively. Thus, we can write
\begin{equation*}
\sum_{i=1}^{M_r}{\frac{T_p^{\text{final}} \varepsilon_{i,r}}{\varepsilon_{i,r}-\varepsilon_1}-{\frac{T_r \varepsilon_{i,r}}{\varepsilon_{i,r}-\mu_r}}} =\varepsilon_1 \sum_{i=1}^{M_r}{{\frac{T_p^{\text{final}}}{\varepsilon_{i,r}-\varepsilon_1}}-{\frac{T_r}{\varepsilon_{i,r}-\mu_r}}},
\end{equation*}
then obtain Eq.~\ref{TEquil}.
Based on the predicted Onsager matrix $\hat{L}$ from Eq.~\ref{OnsagerFromCurrents} and transport equation Eq.~\ref{TwoTerminal}, we rewrite
$J_u-\varepsilon_1 J_\rho=0$ as
\begin{equation*}
\left(L_{uu}F_u+L_{u\rho}F_{\rho}\right)-\varepsilon_1 \left(L_{\rho u}F_u+L_{\rho\rho}F_{\rho}\right)=0,
\end{equation*}
where two forces are $F_u=\frac{1}{T_r}-\frac{1}{T_p^{\text{final}}}$ and $F_{\rho}=\frac{\varepsilon_1}{T_p^{\text{final}}} - \frac{\mu_r}{T_r}$.
Hence, we obtain Eq.~\ref{TfinalL}.

When the probe leaks $\alpha$ power each round, a non-equilibrium steady state can be achieved between the probe and the reservoir in a finite round number, even if the reservoir has a forbidden chemical potential. In this case, the power gain provided by the reservoir to the probe equals the power leaked in the probe $\Delta P=\int_{0}^{z_{\text{end}}}{J_\rho dz}=\alpha P_p$. As a result, the temperature and chemical potential of the steady state in the probe satisfy
\begin{equation}\label{forTalpha}
\begin{split}
\left(1-\alpha\right)P_r &=\sum_{i=1}^{M_r}{\frac{T_p^{s}}{\varepsilon_{i,r}-\mu_p^{s}}}+\alpha\sum_{i=1}^{M_p}{\frac{ T_p^{s} }{\varepsilon_{i,p}-\mu_p^{s}}},\\
\left(1-\alpha\right)U_r &=\sum_{i=1}^{M_r}{\frac{\varepsilon_{i,r}T_p^{s}}{\varepsilon_{i,r}-\mu_p^{s}}}+\alpha\sum_{i=1}^{M_p}{\frac{\varepsilon_{i,p} T_p^{s} }{\varepsilon_{i,p}-\mu_p^{s}}}.
\end{split}
\end{equation}
where $P_r=\sum_{i=1}^{M_r}{\frac{T_r}{\varepsilon_{i,r}-\mu_r}}$ and $U_r=\sum_{i=1}^{M_r}{\frac{\varepsilon_{i,r} T_r}{\varepsilon_{i,r}-\mu_r}}$ are input power and internal energy in the reservoir. From Eq.~\ref{forTalpha}, the chemical potential of steady state can be can be represented as $\mu_p^s=\frac{(1-\alpha)U_r-(\alpha M_p+M_r)T_p^s}{(1-\alpha)P_r}$. Hence, we obtain Eq.~\ref{Talpha} from Eq.~\ref{forTalpha} and one
can solve Eq.~\ref{Talpha} using the method described in Ref.~\cite{parto2019thermodynamic}.

\section{Acknowledgments}
\begin{acknowledgments}
National Natural Science Foundation of China (Grant No. 12274161) and the Innovation Project of Optics Valley Laboratory. J.~H.~J. thanks the National Natural Science Foundation of China (Grant No. 12125504), the ``Hundred Talents Program'' of the Chinese Academy of Sciences, and the Gusu leading scientist program of Suzhou city.
\end{acknowledgments}
 
\nocite{*}

\bibliography{apssamp}

\begin{figure}[!ht]
\centering
\includegraphics[width=0.5\textwidth]
{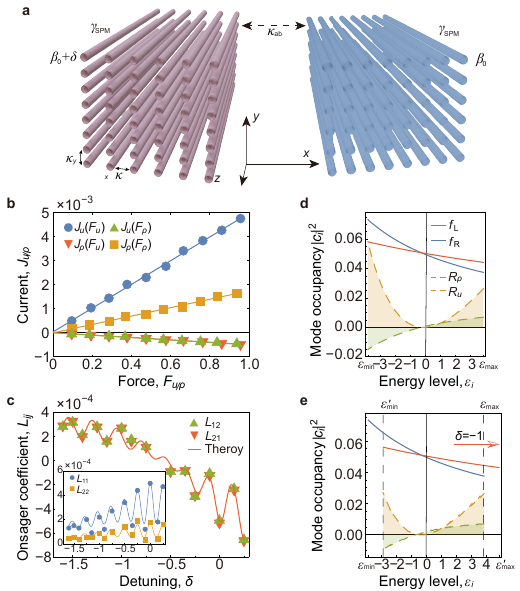}
\captionsetup{font={stretch=1.25},justification=raggedright,singlelinecheck=false}
\caption{\label{Onsager_Current}\textbf{Demonstration of Onsager’s theory and the interpretation of inverse current.} \textbf{a}, A schematic of coupled nonlinear waveguide lattices. Two lattices are evanescently coupled in the x direction, $\kappa_{ab}$ represents the coupling strength. $\beta_0$ and $\gamma_{SPM}$ represent the propagation constant and the nonlinear coefficient of self-phase modulation in a single waveguide, respectively. Each waveguide in the lattice is evanescently coupled to the auxiliary waveguides and $\kappa_x$/$\kappa_y$ represents the coupling strength in x/y direction. 
\textbf{b}, The simulation results demonstrate the validity of Onsager's linear irreversible theory in weakly nonlinear waveguide lattice. The markers represent the simulation results, and the solid lines represent their linear fitting. 
Two lattices are the same $20\times 20$ array, where $(\kappa_x,\kappa_y)=(0.9,1.1)$, $\gamma_{SPM}=1$ and $\kappa_{ab}= 0.1$. 
\textbf{c}, The Onsager coefficients ($L_{ij}$) vary with the detuning (spectrum shift), $\delta$, of the left subsystem. The solid lines are predicted from Eq.~\ref{Practice} with $\sigma=0.06$, magnified 7.5 times.
\textbf{d}, The R-J distributions (solid lines) and R-J factors (dashed line) of two subsystems with forces $F_u=1$, $F_\rho=0.3$ and detuning $\delta=0$. Due to $L_{\rho u}<0$ (as shown in \textbf{c}), $J_\rho$ should oppose the two forces $F_u$ and $F_\rho$. Correspondingly, the shaded area of the positive R-J factor of photon current ($R_\rho>0$) is smaller than that of the negative part ($R_\rho<0$), revealing a negative photon current.
\textbf{e}, Spectrum shifting $\delta=-1$ reverses current direction. Due to $L_{21}>0$ (as shown in \textbf{c}), $J_\rho$ should have the same direction as the two forces. The shaded area of R-J factor $R_\rho$ also reveals a positive photon current. All simulation results are averaged in 3000 ensemble copies.}
\end{figure}

\end{document}